\documentclass[a4paper,11pt]{article}
\usepackage{pos}

\usepackage{csquotes}        
\usepackage{hyperref}          
\usepackage{placeins}          

\newcommand{\tcr}[1]{\textcolor{black}{#1}}

\newcommand{\rb}[1]{\left(#1\right)}
\newcommand{\sq}[1]{\left[#1\right]}
\newcommand{\expe}[1]{\text{e}^{#1}}

\DeclareRobustCommand{\eqnr}[1]{Eq.~$\left(\ref{#1}\right)$}

\newcommand{\Fig}[1]{Figure \ref{#1}}
\newcommand{\Tab}[1]{Table \ref{#1}}

\newcommand{\Figtwo}[2]{Figures \ref{#1} and \ref{#2}}

\newcommand{\Refl}[1]{Ref.~\cite{#1}}    
\newcommand{\Refltwo}[2]{Refs.~\cite{#1,#2}}    

\newcommand{\mpiv}{239(1)}
\newcommand{\mcR}{\mathcal{R}}

\title{Charm baryons at finite temperature on anisotropic lattices}

\author*[a]{Ryan Bignell}
\author[a,b,c]{Gert Aarts}
\author[a]{Chris Allton}
\author[a]{M.~Naeem Anwar}
\author[a]{Timothy J.~Burns}
\author[d]{Benjamin J\"ager}

\affiliation[a]{Department of Physics, Swansea University, Swansea, SA2 8PP, United Kingdom}
\affiliation[b]{European Centre for Theoretical Studies in Nuclear Physics and Related Areas (ECT*)}
\affiliation[c]{Fondazione Bruno Kessler, Strada delle Tabarelle 286, 38123 Villazzano (TN), Italy}
\affiliation[d]{CP3-Origins \& Danish IAS, Department of Mathematics and Computer Science, University of Southern Denmark, 5230, Odense M, Denmark}

\abstract{Singly, doubly and triply charmed baryons are investigated at multiple temperatures using the anisotropic \textsc{FASTSUM} \enquote{Generation 2L} ensembles. We discuss the temperature dependence of these baryons’ spectra in both parity channels with a focus on the confining phase. To further qualify the behaviour of these states around the pseudocritical temperature, we investigate the effect of chiral symmetry restoration for light quarks. We find that an estimate of the pseudocritical temperature can still be found from positive and negative-parity charmed baryon correlators, even when parity doubling itself is not very evident (as expected).}

\FullConference{%
  The 39th International Symposium on Lattice Field Theory (Lattice2022),\\
  8-13 August, 2022 \\
  Bonn, Germany 
}

\begin{document}
\maketitle

\section{Introduction}
As the temperature increases, the confining hadronic phase of Quantum Chromodynamics (QCD) smoothly~\cite{Aoki:2006we} deforms to a deconfining quark-gluon plasma (QGP). In this phase the lighter quarks are deconfined while the heavier charm and bottom quarks may remain bound until higher temperature. This transition is experimentally investigated in particle collider experiments such as the Large Hadron Collider~\cite{Armesto:2015ioy,Busza:2018rrf,Berges:2020fwq} and the Relativistic Heavy Ion Collider~\cite{Braun-Munzinger:2015hba}.
\par
Here we examine the charmed baryon spectrum through lattice QCD simulations. We vary the temperature by use of a series of anisotropic ensembles with different temporal extents. Anisotropic ensembles allow fine control of temperature and allow for a greater number of temporal data points at each temperature which is key for spectral studies. Charm baryons are of interest as the heavier charm quark is produced in relatively high proportions in heavy-ion collisions and the QGP exists long enough for the charm quarks to diffuse~\cite{Yao:2018zze,Begel:2022kwp} and contribute to the hadronisation process.
\par
In particular we examine the masses of the states as a function of temperature. The extraction of baryon masses at high temperature becomes difficult due to both the reduced number of temporal points and the more fundamental question of whether the hadron is a bound state at these temperatures. The extracted baryon masses may be of interest to phenomenology~\cite{Yao:2018zze}. 
\par
As extracting baryon masses becomes difficult at the highest temperatures, we instead compare positive and negative parity channels~\cite{Aarts:2017rrl,Aarts:2018glk,Aarts:2020vyb} using the correlation functions directly, to investigate the effect of the restoration of chiral symmetry in the QGP. While parity doubling is not expected due to the large charm quark mass, we still note a clear change of behaviour, occurring around the thermal crossover temperature for singly and doubly charmed baryons.
\FloatBarrier
\section{Lattice setup}
\begin{table}[]
  \centering
    \caption{The \textsc{FASTSUM} Generation 2L ensembles used in this work which have lattice size $32^3 \times N_\tau$ and temperature $T = 1/\rb{a_\tau N_\tau}$. The spatial lattice spacing is $a_s = 0.1121\rb{3}$ fm, renormalised anisotropy $\xi = a_s/a_\tau = 3.453(6)$ and the pion mass $m_\pi = \mpiv$ MeV~\cite{Wilson:2019wfr}. We use $\sim 1000$ configurations and eight (random) sources for a total of $\sim 8000$ \tcr{sources} at each temperature. Full details of these ensembles may be found in \Refl{Aarts:2020vyb,Aarts:2022krz}.}
  \begin{tabular}{r|rrrrr||rrrrrr}
  $N_\tau$ & 128 & 64 & 56 & 48 & 40 & 36 & 32 & 28 & 24 & 20 & 16\\ \hline
  $T\,\, \rb{\text{MeV}}$ & 47 & 95 & 109 & 127 & 152 & 169 & 190 & 217 & 253 & 304 & 380 \\
\end{tabular}%
\label{tab:ensembles}
\end{table}
We calculate two-point correlation functions using standard baryon interpolating operators~\cite{Leinweber:2004it,Edwards:2004sx} on the \textsc{FASTSUM} \enquote{Generation 2L} thermal ensembles~\cite{Aarts:2020vyb,Aarts:2022krz}. These span a wide range of temperatures using a fixed-scale anisotropic approach with a Wilson-clover fermion action and Symanzik-improved gauge action~\cite{Edwards:2008ja, Sheikholeslami:1985ij,Symanzik:1983dc,Symanzik:1983gh}. The pion mass is $239(1)$ MeV. \Tab{tab:ensembles} shows the details of these ensembles.
\par
Excited state effects are reduced by the use of Gaussian smearing~\cite{Gusken:1989qx} at the source and the sink. The root-mean-square radius of the smearing profile is $\sim 6.1$ lattice sites, an amount chosen such that the ground-state isolation for the zero temperature positive parity nucleon is good.
\section{Masses}
The zero-momentum projected two-point correlation function has contributions from multiple positive and negative parity energy eigenstates. Accordingly, we fit the correlator using functions of the form
\begin{align}
  G\rb{\tau} = \sum_{\alpha=1}^{N}\, A_\alpha\expe{-a_\tau E^{+}_\alpha\tau/a_\tau} + B_\alpha\expe{-a_\tau E_\alpha^{-}\rb{N_\tau - \tau/a_\tau}},
  \label{eqn:expFit}
\end{align}
where the number of exponentials is allowed to vary $N=1,2,3$; $E_\alpha^{+}$ are the positive parity energies and $E_\alpha^{-}$ are the negative parity energies. $N$ is set by examination of the Gaussian Bayes factor~\cite{Lepage:2001ym,peter_lepage_2021_5777652} which tells us that additional exponential terms would not increase the quality of the fit.
\par
\begin{figure}[tb]
  \centering
  \includegraphics[width=0.9\columnwidth]{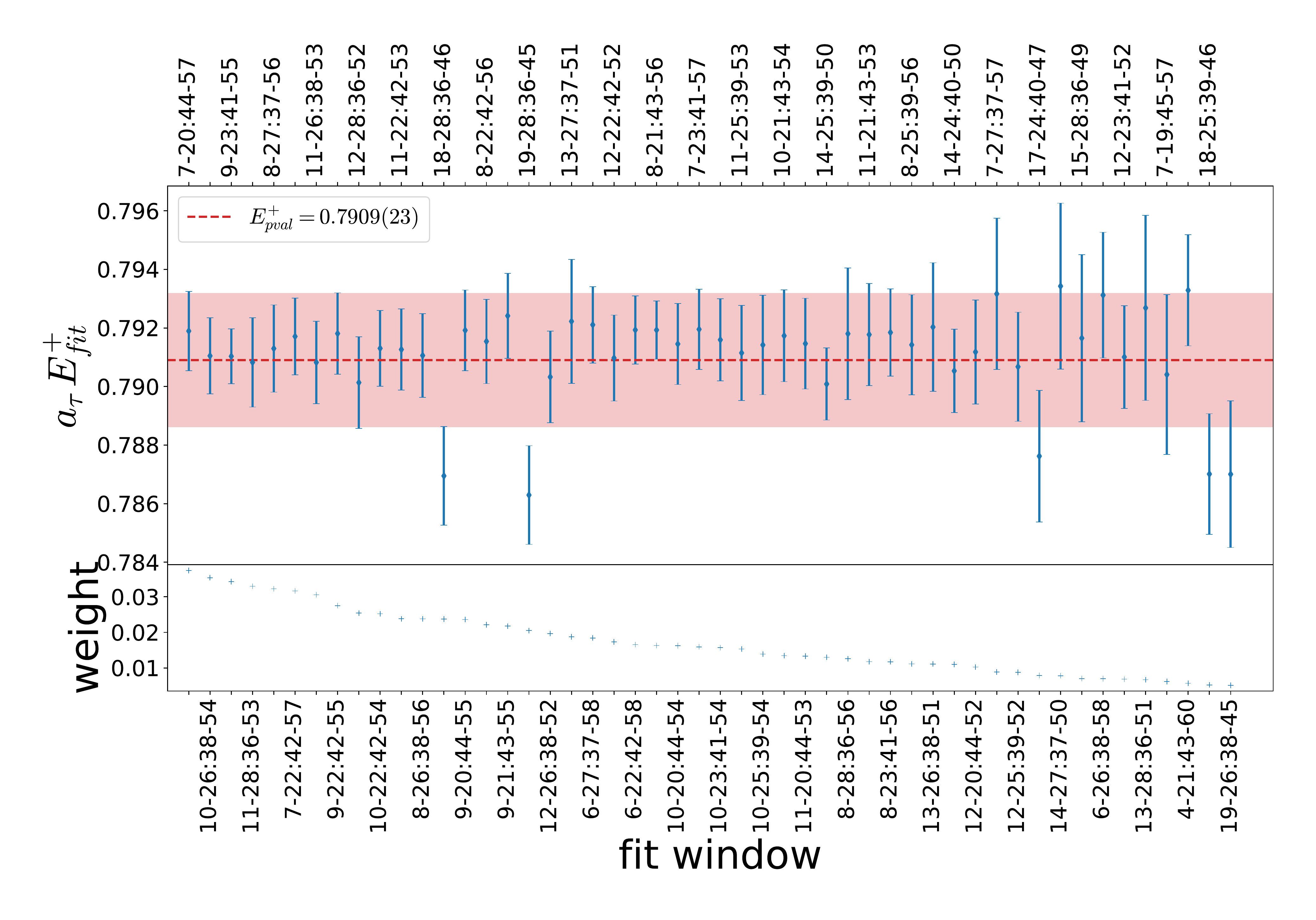}%
  \caption{\label{fig:Weights}Highest weighted fit results for the $N_\tau = 64$ ${}^{\frac{3}{2}}\Omega_{ccc}$ positive parity energy as a function of fit window. The weights of \Refl{Rinaldi:2019thf} are also shown, as is the resulting model averaged fit value. Both sides of the correlator are fit symmetrically e.g. \enquote{10-26:38-54} fits points $[10,26]$ and $[38,54]$ simultaneously.}
  \end{figure}
\begin{figure}[tb]
  \centering
  \includegraphics[width=0.8\columnwidth]{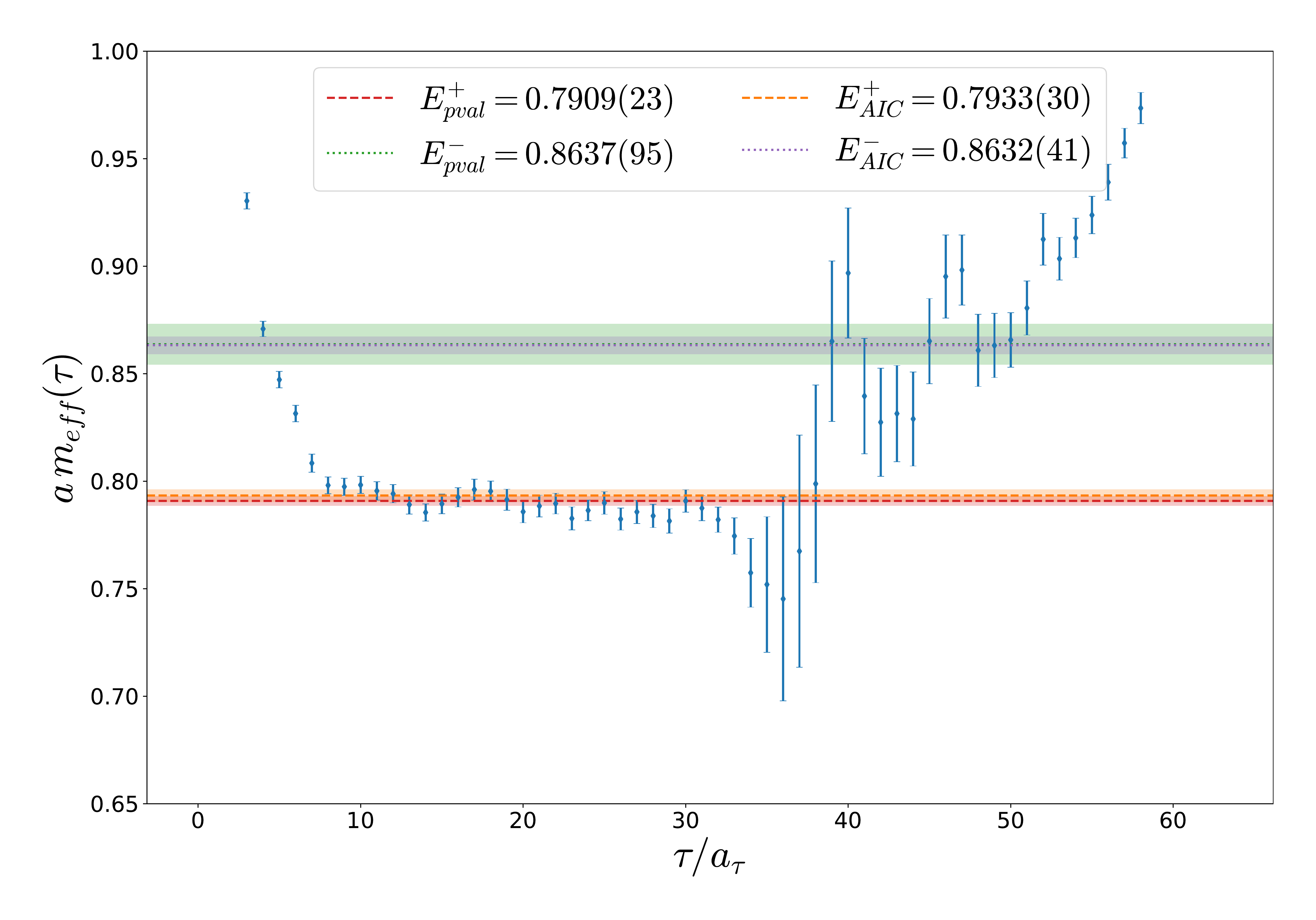}%
  \caption{\label{fig:effMass}Effective mass for the $N_\tau = 64$ ${}^{\frac{3}{2}}\Omega_{ccc}$ state showing the results of the two model averaging methods described in \Refltwo{Rinaldi:2019thf}{Jay:2020jkz}.}
  \end{figure}
In order to ensure a robust determination of the masses, we make use of the model averaging methods presented in \Refltwo{Rinaldi:2019thf}{Jay:2020jkz}. Here all possible temporal subsets of the correlator (\enquote{fit-windows}) are considered and the resulting fits are averaged with some weight. Outlying fits are excluded via a consideration of the 30\% of fits with the highest weight from the Akaike information criterion weights~\cite{Jay:2020jkz}. The final mass is produced by picking the averaged result which most clearly represents the underlying fits. An example of this is shown in \Figtwo{fig:Weights}{fig:effMass} for the positive parity $N_\tau = 64$ ${}^{\frac{3}{2}}\Omega_{ccc}$ state. In \Fig{fig:Weights} we show the highest weighted fits using the \enquote{p-value} method~\cite{Rinaldi:2019thf} where the fit-window is shown above and below. As we fit both sides of the correlator, the label reflects this. The averaged result encompasses the clear majority of fit weights and is not an outlier or unstable result. A systematic uncertainty from the choice of averaging method is also considered.\phantom{\cite{HadronSpectrum:2012gic,Padmanath:2021lzf}}
\begin{figure}[tb]
  \centering
  \includegraphics[width=0.9\columnwidth]{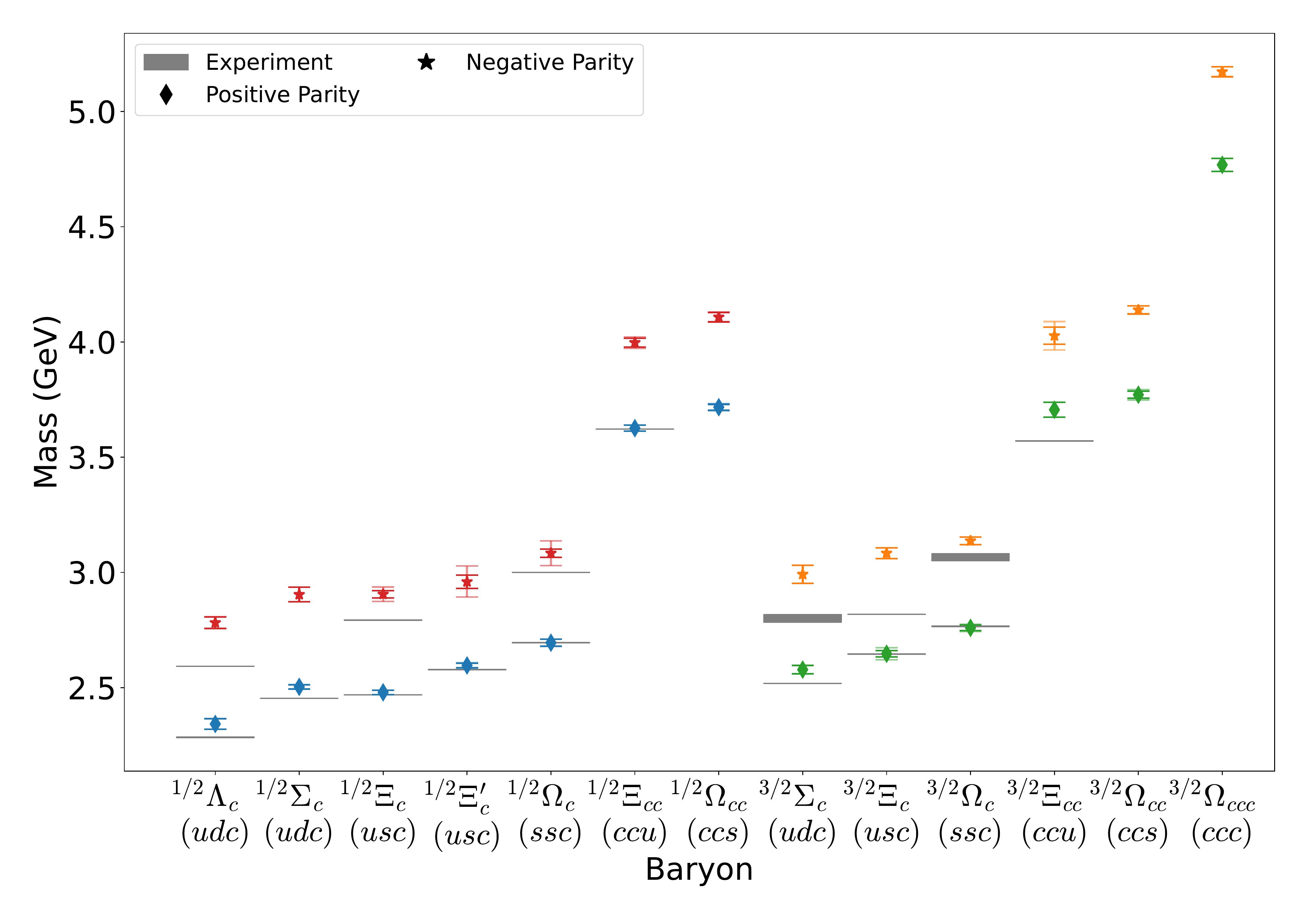}%
  \caption{\label{fig:spectAll}The ground-state charm baryon spectrum from the $N_t = 128$ zero-temperature ensemble. Positive parity states are shown by \enquote{diamonds} $\rb{\blacklozenge}$ and the negative parity by \enquote{stars} $\rb{\star}$. Inner error bars represent the statistical uncertainty and the outer incorporates a systematic from the choice of averaging method. The grey bands represent experimental results from the Particle Data Group~\cite{Zyla:2020zbs} with uncertainties from different charge states added in quadrature. For the $\Omega_{c}$ we use the spin-parity assignments suggested in \Refltwo{Cheng:2021qpd}{Chen:2022asf}. We select $\Omega_{c}^{3/2-} = \Omega_{c}(3065)$ and include a systematic uncertainty encompassing the $\Omega_{c}(3050)$ mass due to the former's better known width.}
\end{figure}
\par
A description of the charm baryon states investigated in this study is provided in \Fig{fig:spectAll} which displays the ground-state charm baryon spectrum on the \enquote{zero temperature} $N_\tau=128$ ensemble. The zero-temperature results are in good agreement with the experimental measurements for the states containing only $s$ or $c$ quarks as these quarks have been tuned to their physical values~\cite{HadronSpectrum:2012gic}. States containing $u$ or $d$ quarks are heavier than nature which is expected on these ensembles. The emphasis in this work is not on extreme precision zero-temperature spectroscopy~\cite{HadronSpectrum:2012gic,Padmanath:2021lzf} and so \Fig{fig:spectAll} is only a guide to the charmed baryons considered in this work.
\begin{figure}[tb]
  \centering
  \includegraphics[width=0.48\columnwidth]{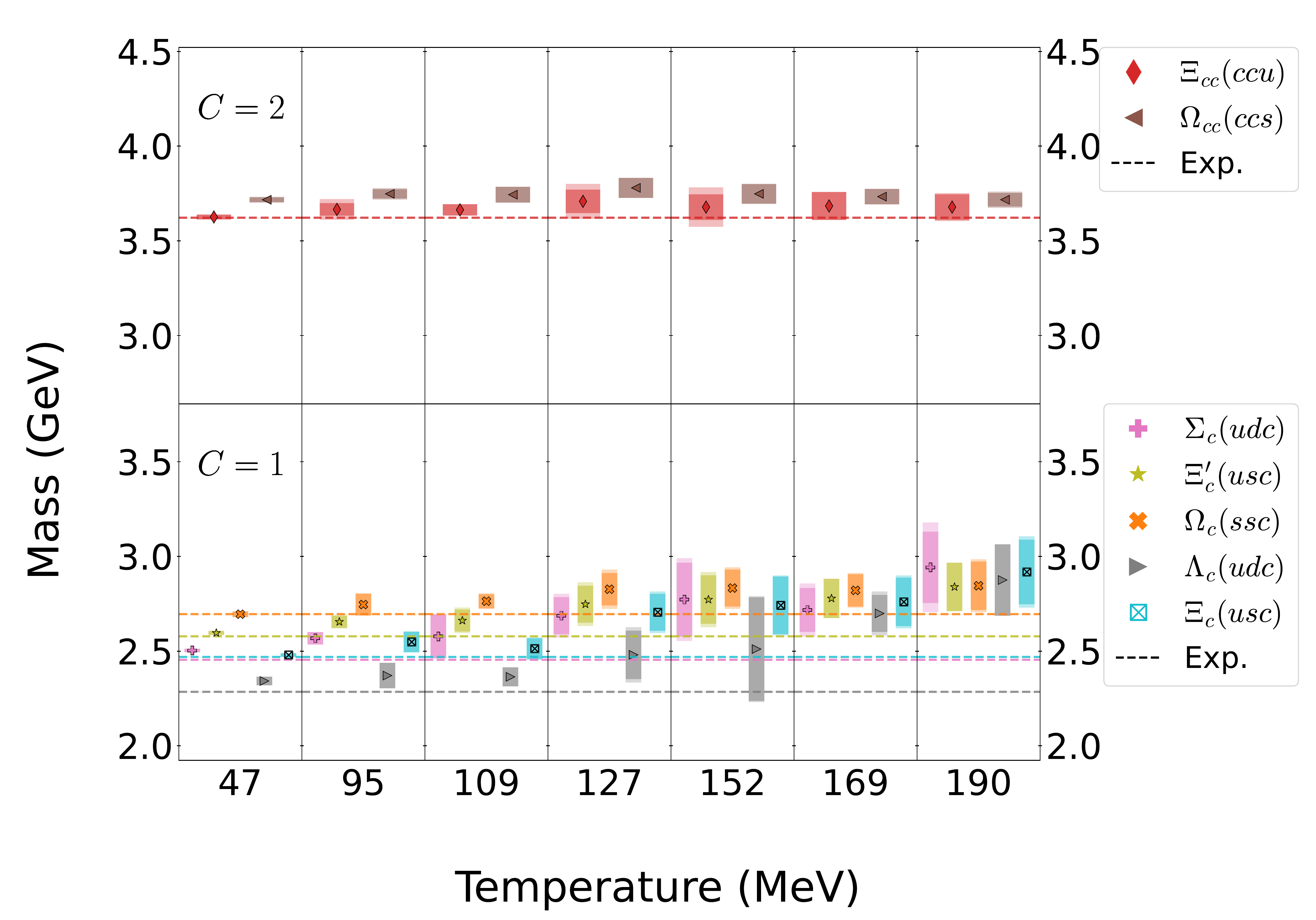}%
  \includegraphics[width=0.48\columnwidth]{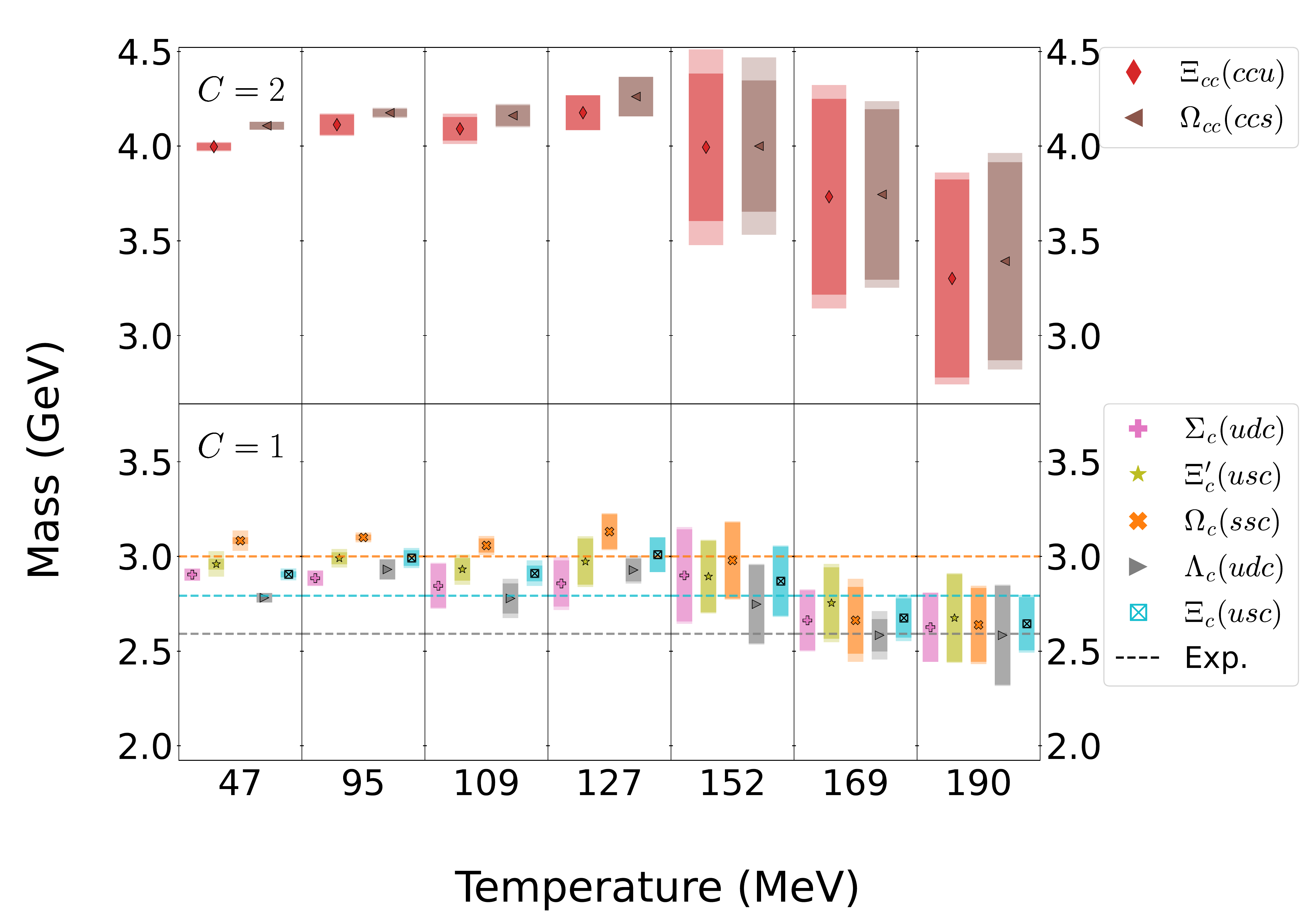}%
    \caption{\label{fig:spectJ1_2P}\textbf{Left:} Mass spectrum of $J^{P} = {1/2}{}^{+}$ baryons as a function of temperature. Dashed lines are zero-temperature experimental results~\cite{Zyla:2020zbs} to guide the eye. The inner shaded region represents the statistical uncertainty, and the outer incorporates the systematic from the choice of averaging method. The same vertical scale is used in each subplot, making comparison of the uncertainties possible. Some states have increasing uncertainties around and past $T_{pc}$. This may indicate these states becoming unbound at these temperatures. \textbf{Right:} Mass spectrum of $J^{P} = {1/2}{}^{-}$ baryons.}
\end{figure}
\par
We extend the analysis of \Fig{fig:spectAll} to non-zero temperature for the $J=1/2$ charmed baryons in \Fig{fig:spectJ1_2P}. Here the positive and negative parity sectors are shown separately, enabling an examination of the temperature dependence. It is clear that within each parity sector, different channels display qualitatively similar behaviour. \Fig{fig:spectJ1_2P} suggests that the effect of temperature decreases with charmness $C$.
\par
\begin{figure}[tb]
  \centering
  \includegraphics[width=0.48\columnwidth]{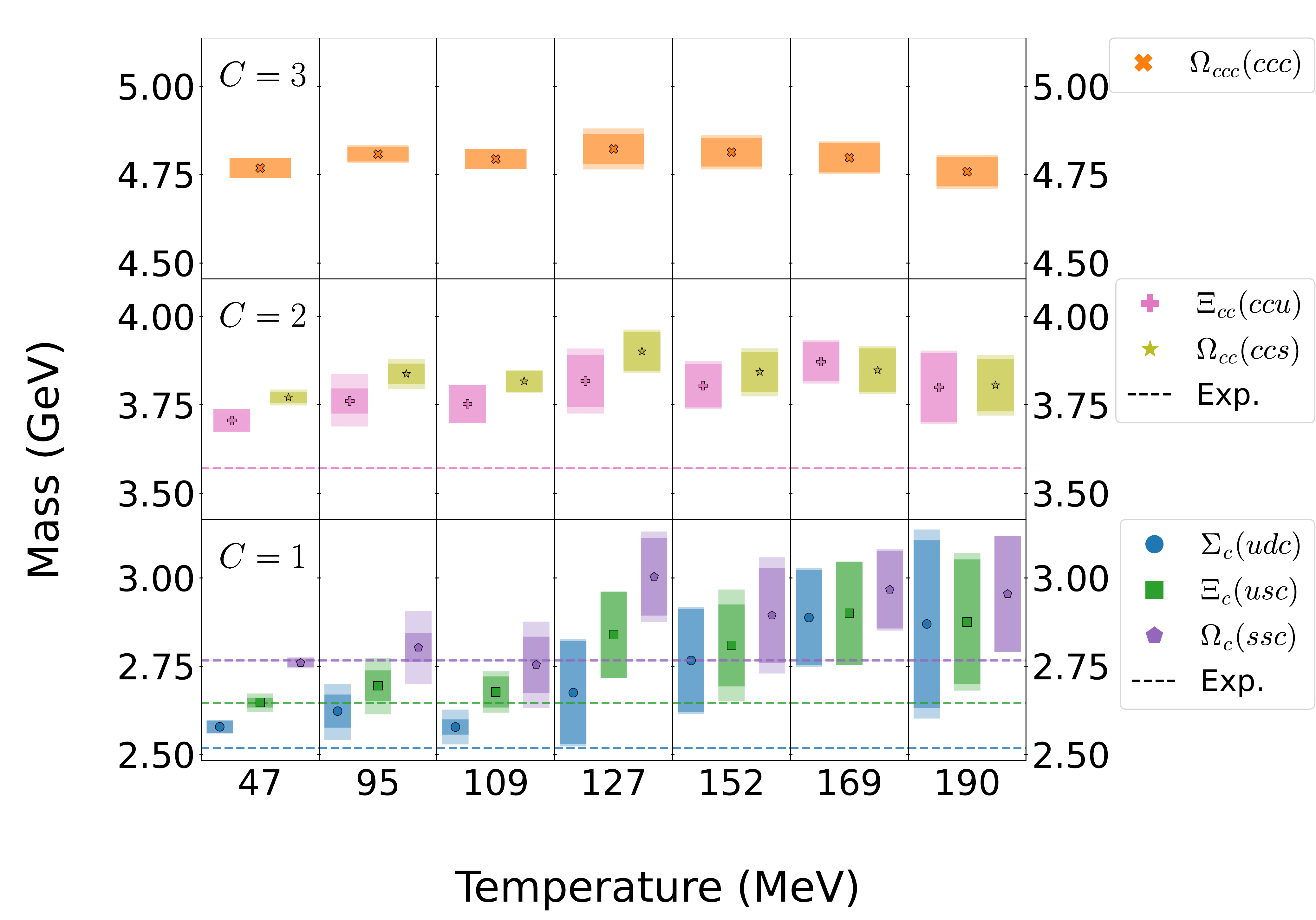}%
  \includegraphics[width=0.48\columnwidth]{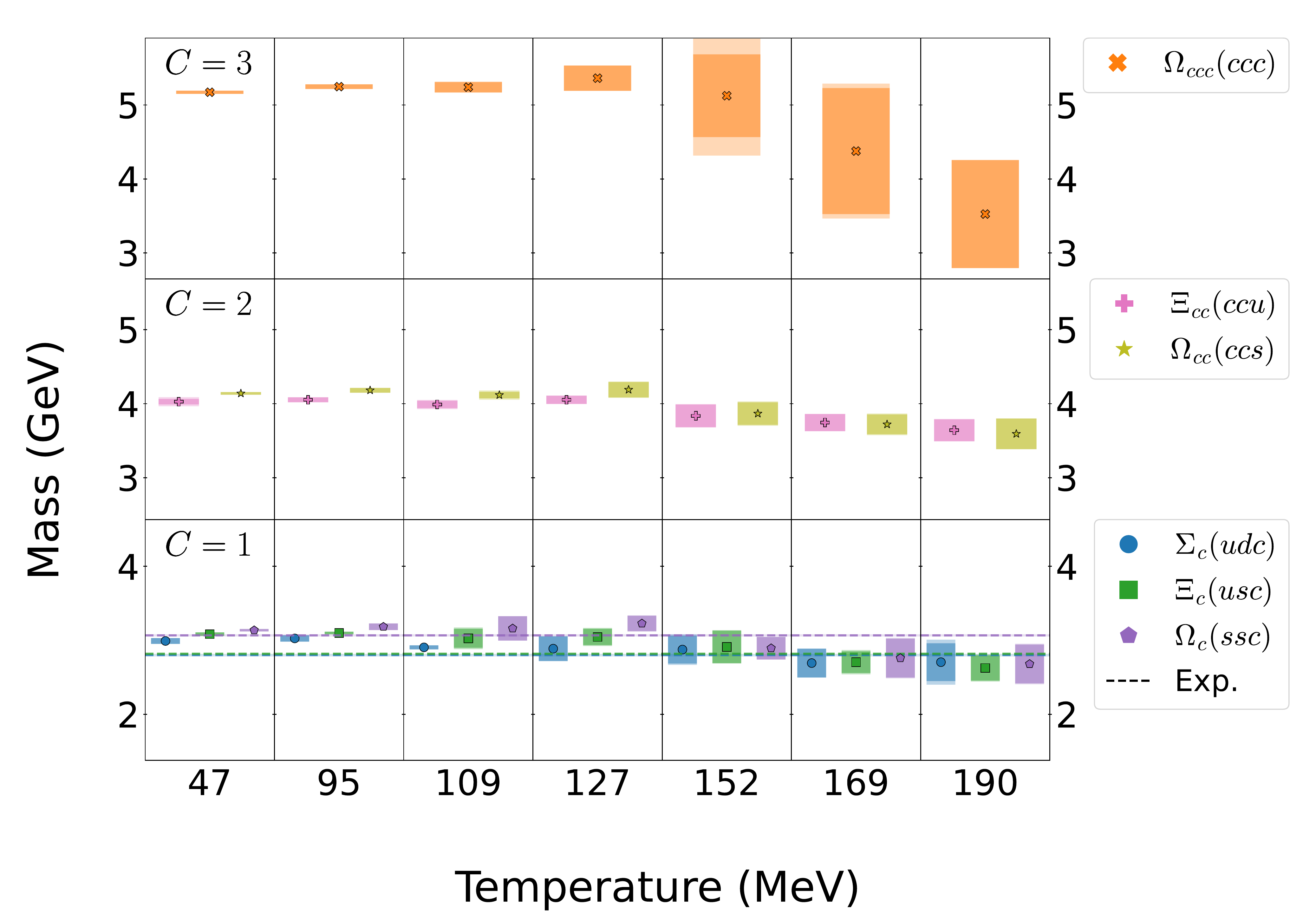}%
  \caption{\label{fig:spectJ3_2P}\textbf{Left:} Mass spectrum of $J^{P} = {3/2}{}^{+}$ baryons as a function of temperature. Details as in \Fig{fig:spectJ1_2P}. \textbf{Right:} Mass spectrum of $J^{P} = {3/2}{}^{-}$ baryons. Here positive and negative parity plots have different vertical scales.}
\end{figure}
In \Fig{fig:spectJ3_2P} a similar picture emerges for the $J=3/2$ baryons. In this figure, the positive and negative parity plots do not share the same vertical scale however each subplot is the same. Here however the positive parity, $C=3$ $\Omega_{ccc}$ mass is unaffected by temperature within uncertainties. This agrees with the pattern observed for the $J=1/2$ sector where the $C=1$ states displayed greater temperature dependence than the $C=2$.
\par
Determining the mass at non-zero temperature is a difficult task. This can be observed by the increasing uncertainty present in the preceding figures. The extracted masses suggest that above the transition temperature of $\sim 167$ MeV~\cite{Aarts:2019hrg} the positive and negative parity sectors show greater temperature dependence. \tcr{The suitability of the exponential ansatz of \eqnr{eqn:expFit} at high temperatures is an interesting problem. Already at $N_\tau = 32$ ($T=190$ MeV) the exponential fits do not seem to be describing all the states well. In particular the $\Xi_{cc}$ is well described at this temperature, but not at the next hottest. The lighter states are ill-behaved at lower temperatures, which may be an indication they are no longer bound states at those temperatures. Nevertheless, we show the obtained masses, including the large uncertainty produced by our analysis - a sign of the breakdown of the ansatz. This problem could be better investigated with e.g.~a spectral function function analysis or a correlator with more ground-state isolation before the interference of the positive and negative parity terms in the centre of the lattice. Such isolation could be achieved via a better determination of the correlator on these ensembles, or an ensemble with more temporal points at the same temperature. Both these methods are under investigation.}
\par
In summary, our results suggest that the temperature dependence is greater for negative parity states and in the $J=3/2$ sectors.
\section{Parity Doubling}
Chiral symmetry is expected to-be restored in the QGP. For light and strange baryons, this results in a parity-doubling signal at the level of the correlators~\cite{Aarts:2017rrl,Aarts:2018glk}. For charmed baryons for the temperatures studied in this work, such a signal is not expected, since the charm quark mass breaks chiral symmetry explicitly. Nevertheless, it is interesting to study whether restoration of chiral symmetry for lighter quarks leads to visible effects in the charmed baryon sector.
\par
We therefore turn to the parity doubling $R$-quantity considered in \Refltwo{Datta:2012fz}{Aarts:2015mma}. This method has the advantage that no fits are required, enabling high statistical precision. We form the quantities
\begin{align}
  \mcR\rb{\tau} &= \frac{G^{+}\rb{\tau} -G^{-}\rb{\tau}}{G^{+}\rb{\tau} + G^{-}\rb{\tau}}, \label{eqn:mcRratio} \\
  R\rb{n_0} &= \frac{\sum_{n=n_0}^{\frac{1}{2}N_\tau -1}\, \mcR\rb{\tau_{n}}/\sigma_{\mcR}^2\rb{\tau_{n}}}{\sum_{n=n_0}^{\frac{1}{2}N_\tau -1}\,1/\sigma_{\mcR}^2\rb{\tau_{n}}},
  \label{eqn:Rratio}
\end{align}
where $G\rb{N_\tau a_t - \tau} = G^{-}\rb{\tau}$ is the negative parity correlator, $G\rb{\tau} = G^{+}\rb{\tau}$ the positive parity correlator, and $\sigma_{\mcR}\rb{\tau_{n}}$ denotes the statistical error for $\mcR\rb{\tau_{n}}$. The sum over time slices $\tau_{n}\in \sq{n_0=4, N_{\tau} / 2 - 1}$ is chosen such that excited state contributions and lattice artefacts at small $\tau_{n}$ are suppressed. Small shifts in $n_0$ do not have a qualitative effect on the results.
\begin{figure}[tb]
  \centering
  \includegraphics[width=0.48\columnwidth]{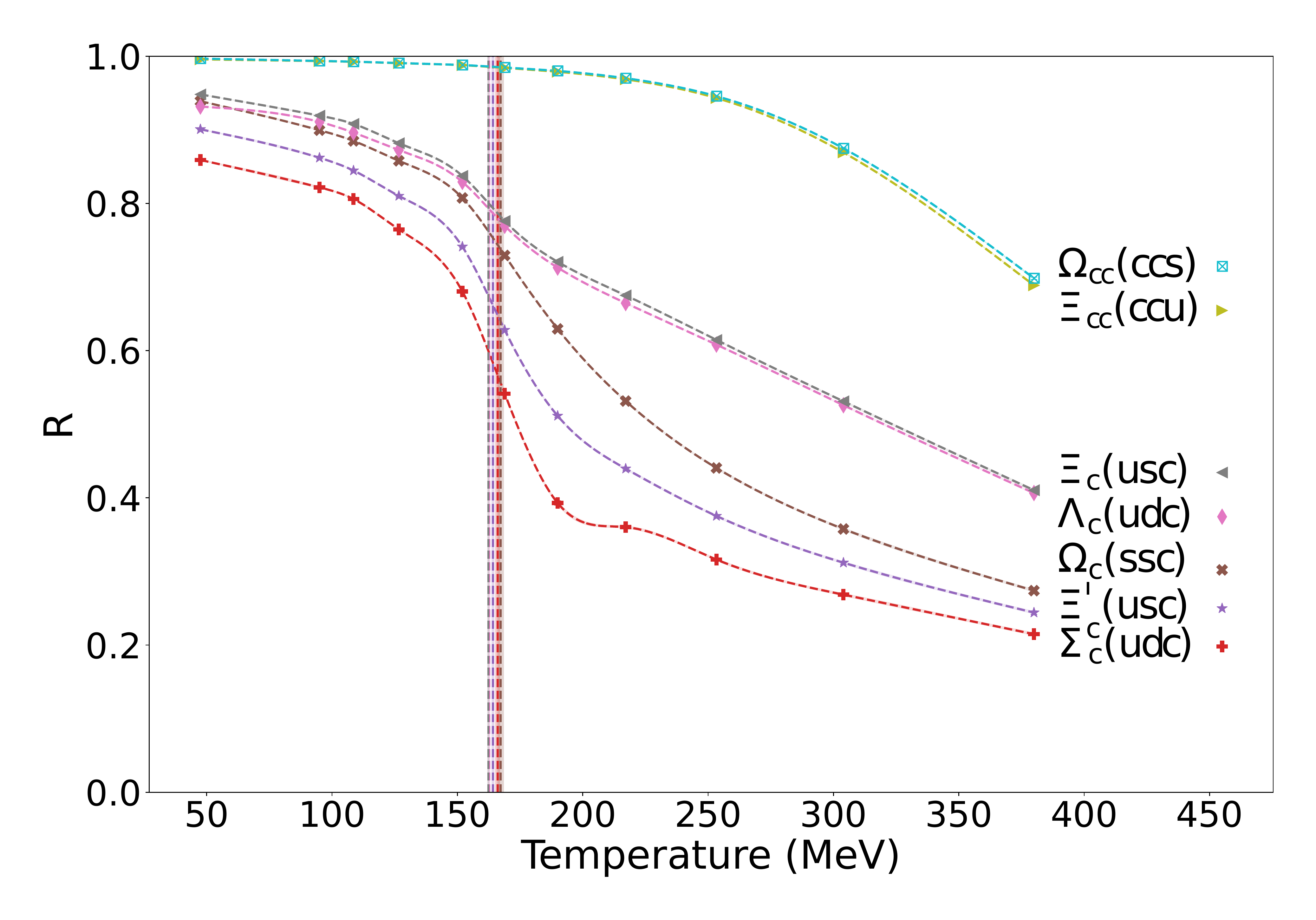}%
  \includegraphics[width=0.48\columnwidth]{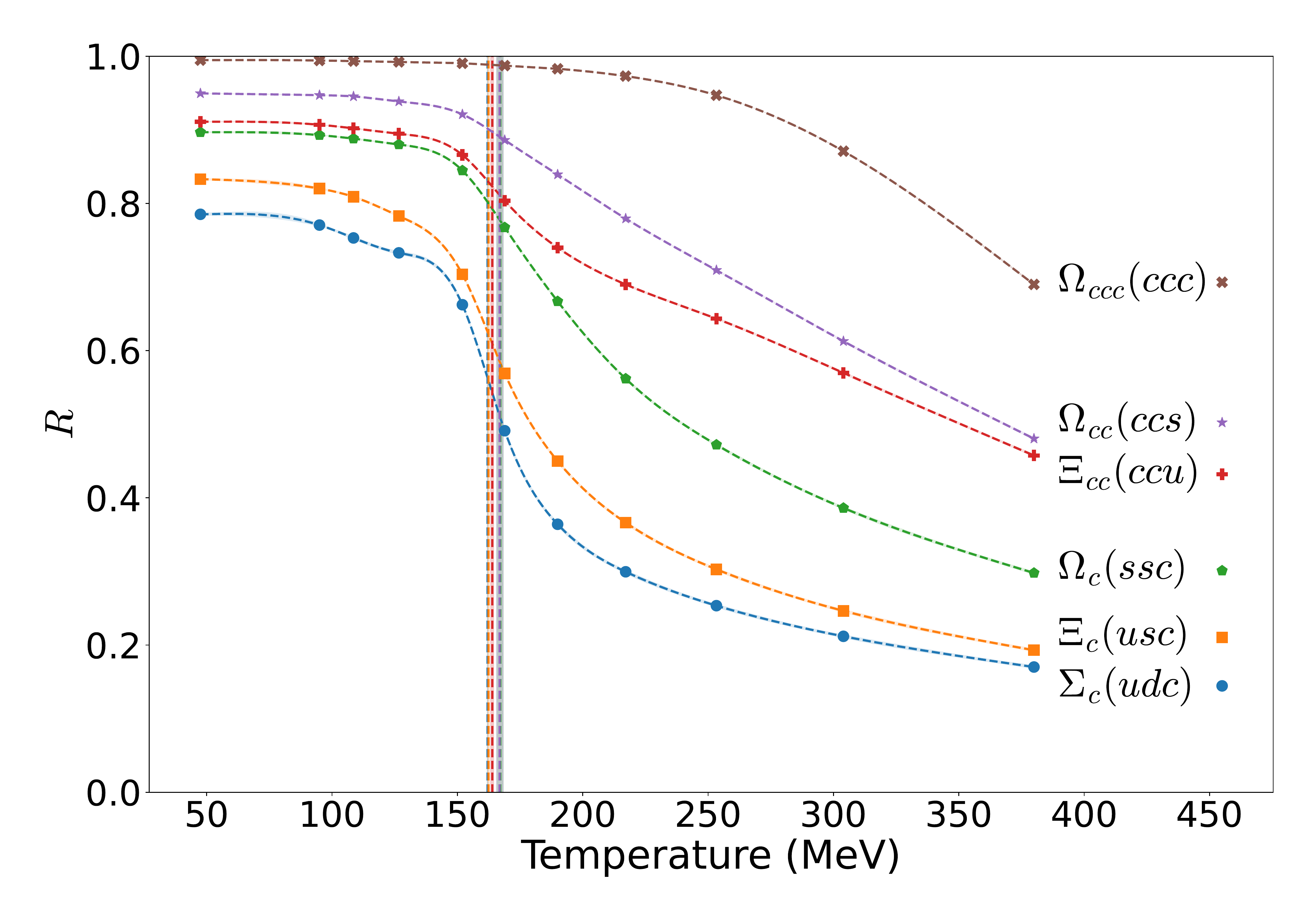}%
  \caption{\label{fig:1/2RRatio}\textbf{Left:} $R$-quantity of \eqnr{eqn:Rratio} for $J=\frac{1}{2}$ baryons. The lines are cubic splines used to find the transition temperature which are shown in the vertical bands. \textbf{Right:} $R$-quantity of \eqnr{eqn:Rratio} for $J=\frac{3}{2}$ baryons.}
\end{figure}
\par
As is clear from \eqnr{eqn:mcRratio}, $R\rightarrow 0$ when positive and negative correlators become degenerate, which coincides with chiral symmetry restoration. Conversely, when the states are non-degenerate with $m_{+} \ll m_{-}$, $R$ will be close to one. This is indeed the behaviour we see in \Fig{fig:1/2RRatio}.
\par
Due to the heavier mass of the charm quarks, chiral symmetry restoration is not expected, even at our highest temperatures. Again this is the effect seen in \Fig{fig:1/2RRatio} with it being particularly notable that the $R$-quantity increases monotonically with the number of charm quarks.
\par
We also observe an intriguing similarity in the behaviour of the $R$-quantity for states (of a given $J$) which have the same charm content, and which belong to the same $SU(3)$ flavour multiplets, namely:
\begin{align*}
  C=+2,\,\quad SU(3)\,\, \mathbf 3:\,\quad &\Omega_{cc}\rb{ccs}, \,\Xi_{cc}\rb{ccu} \\
  C=+1,\,\quad SU(3)\,\, \mathbf \overline{3}:\, \quad&\Lambda_{c}\rb{udc},\, \Xi_{c}\rb{usc} \\
  C=+1,\, \quad SU(3)\,\, \mathbf 6:\, \quad&\Sigma_{c}\rb{udc}, \, \Xi_{c}^{\prime}\rb{usc},\, \Omega_{c}\rb{ssc}
\end{align*}
\par
A cubic spline interpolation of the data points for each hadron is performed; this enables the inflection point of the curves to be found. For singly-charmed baryons with $J=1/2$ in \Fig{fig:1/2RRatio} left we find the inflection point to be near $T_{pc}$; interestingly, the doubly charmed $J=1/2$ baryons do not have an inflection point. The ability of these inflection points to describe the transition temperature is present even when parity doubling is not nearly manifest, as for the $\Lambda_{c}(udc)$, as seen in \Fig{fig:1/2RRatio} left.
\par
We repeat this exercise for the $J=3/2$ baryons in \Fig{fig:1/2RRatio} right. Here we note that doubly charmed $J=3/2$ baryons display an inflection point, again around $T_{pc}$. This is aligned with the results observed for masses, where the $J=3/2$ states were more affected by temperature than the $J=1/2$ states.
\begin{figure}[tb]
  \includegraphics[width=\columnwidth]{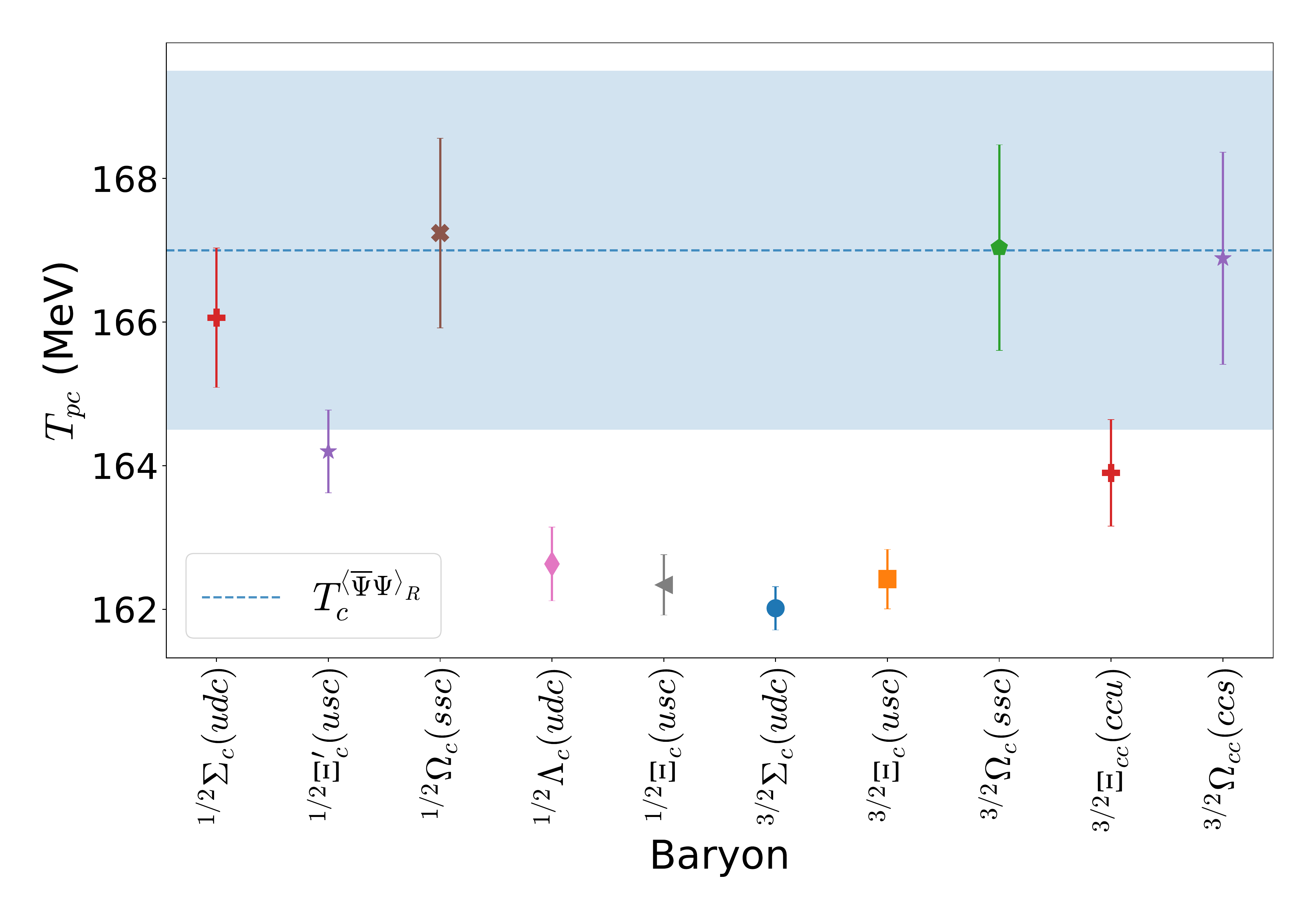}%
  \caption{\label{fig:3/2Inflect}Inflection points of baryon $R$-quantity of \Fig{fig:1/2RRatio}. The blue band is the pseudocritical temperature from the renormalised chiral condensate~\cite{Aarts:2019hrg} adjusted for the change in lattice spacing~\cite{Wilson:2019wfr,Aarts:2022krz}.}
\end{figure}
\par
The inflection points found are presented in \Fig{fig:3/2Inflect} alongside the computation of the pseudocritical temperature as obtained via the inflection point of the renormalised chiral condensate (blue band)~\cite{Aarts:2019hrg}. Excellent precision and good agreement with the previous measurement is seen.
\section{Conclusions \& Future Work}
A variety of singly, doubly and triply charmed baryons have been investigated using lattice QCD. We calculated two-point correlation functions at a range of temperatures using the \textsc{FASTSUM} \enquote{Generation 2L} anisotropic ensembles. Baryon masses were extracted by performing multi-exponential fits to all possible fit windows and considering two different methods with which to weigh the results. In the hadronic phase and just above the crossover, it is possible to extract ground state masses using the method described. At higher temperatures a more sophisticated spectral function analysis would be required.
\par
To investigate the effect of chiral symmetry restoration, the correlator $R$-quantity was considered, to examine the difference between positive and negative parity correlation functions as a function of temperature. Despite the absence of parity doubling (due to the large charm quark mass), a crossover effect can nevertheless be observed, with inflection points close to the pseudocritical temperature.
\par
In future work in preparation, we will examine more sophisticated methods of exploiting the correlators directly, as recently performed for the $D$-mesons in \Refl{Aarts:2022krz}. We anticipate providing the data and analysis tools at that time. In order to improve these results, one could construct a correlator which had more overlap with the ground-state or repeat the analysis on correlators with more temporal points at the same temperature.
\begin{acknowledgments}
This work is supported by STFC grant ST/T000813/1. This work used the DiRAC Extreme Scaling service at the University of Edinburgh, operated by the Edinburgh Parallel Computing Centre and the DiRAC Data Intensive service operated by the University of Leicester IT Services on behalf of the STFC DiRAC HPC Facility (www.dirac.ac.uk). This equipment was funded by BEIS capital funding via STFC capital grants ST/R00238X/1, ST/K000373/1 and ST/R002363/1 and STFC DiRAC Operations grants ST/R001006/1 and ST/R001014/1. DiRAC is part of the UK National e-Infrastructure. We acknowledge the support of the Swansea Academy for Advanced Computing, the Supercomputing Wales project, which is part-funded by the European Regional Development Fund (ERDF) via Welsh Government, and the University of Southern Denmark and ICHEC, Ireland for use of computing facilities. This work was performed using PRACE resources at Cineca (Italy), CEA (France) and Stuttgart (Germany) via grants 2015133079, 2018194714, 2019214714 and 2020214714. M.~N.~A. acknowledges support from The Royal Society Newton International Fellowship.  
\end{acknowledgments}
%
\bibliographystyle{JHEP_arXiv}
\bibliography{skeleton}

\end{document}